# Graph Compact Orthogonal Layout Algorithm*

Kārlis Freivalds and Jans Glagoļevs

Institute of Mathematics and Computer Science
University of Latvia, Raina bulvaris 29, Riga, LV-1459, Latvia
`karlis.freivalds@lumii.lv`, `jansglagolevs@gmail.com`

**Abstract.** There exist many orthogonal graph drawing algorithms that minimize edge crossings or edge bends, however they produce unsatisfactory drawings in many practical cases. In this paper we present a grid-based algorithm for drawing orthogonal graphs with nodes of prescribed size. It distinguishes by creating pleasant and compact drawings in relatively small running time. The main idea is to minimize the total edge length that implicitly minimizes crossings and makes the drawing easy to comprehend. The algorithm is based on combining local and global improvements. Local improvements are moving each node to a new place and swapping of nodes. Global improvement is based on constrained quadratic programming approach that minimizes the total edge length while keeping node relative positions.

## 1 Background

Graph drawing algorithms provide a visually appealing way to present the structure of a graph. Several graph drawing styles are commonly used, each underlining some property of the graph suitable for a particular application. We deal with the orthogonal drawing style where edges are represented by chains of horizontal and vertical line segments connecting the nodes. The goal is to obtain an aesthetically pleasing drawing of a given graph. Common aesthetic criteria include alignment of nodes, small area, few bends and crossings, short edge length. Overlaps of objects are not allowed.

Most of prior work on orthogonal drawing algorithms is dedicated to producing drawings of some provable quality aspect. This is the case of the popular topology-shape-metric approach [5, 9, 17] where the number of bends is minimized respecting some planar embedding. See [6] for an experimental evaluation of these algorithms. There are a number of works achieving proven area bounds, or bounds on the number of bends or both [1–3]. Unfortunately these are worst case bounds and often a given particular graph can be laid out much better as these algorithms produce. Most of the current orthogonal drawing algorithms perform poorly in a practical setting. Even simple heuristics often yield a significant improvement of the drawing quality [8, 16].

We explore the orthogonal drawing problem from a practical point of view where the goal is to produce nice-looking layouts of typical graphs. To achieve this, crossings, bends, area, etc. should be minimized together in some proportion for each particular graph so that the user cannot spot obvious ways of improvement. Such goal is aimed in [3] where the layout process is divided in three phases – node placement, edge routing and port assignment. Unfortunately their node placement phase is weak – each node is placed in a new row and column thus producing large area and long edges. We generally follow this strategy, but implement each phase in a different way.

---

* Supported in part by project No. 2013/0033/2DP/2.1.1.1.0/13/APIA/VIAA/027

In this paper we present an orthogonal layout algorithm which produces good drawings for many practical graphs. At first, the algorithm assigns positions to nodes by putting them on a grid while minimizing the total edge length. Edges are routed afterwards by using standard techniques from integrated circuit layout [14] and minimally adjusting the node placement [10]. Minimizing the total edge length also helps to keep the number of crossings and bends low, although they are not directly minimized. Placing nodes on the grid is an essential ingredient of the algorithm that ensures non-overlapping and nice alignment of nodes characteristic to the orthogonal style.

## 2 Overview of the Algorithm

We consider the drawing model where nodes are represented by rectangles of a given minimum size, edges are represented with orthogonal polylines connecting the associated nodes. Overlaps between nodes or between nodes and edges are not allowed and some minimum distance $\delta$ between them has to be ensured. Only point-wise crossings of edges are allowed (no overlaps of segments of the same direction). We allow nodes to stretch to accommodate adjacent edges but excessive stretching should be minimized. We do not require strict grid placement of nodes and edges but include alignment as an aesthetic criterion to be maximized.

Similarly to [3], the layout process is divided into three phases - node placement, edge routing and normalization (see Figure 1). Node placement is the main phase since it influences the drawing quality the most and other phases depend on it. Then comes edge routing which finds routes for edges that are short and with few bends [14]. Our employed routing algorithm does not minimize crossings, although some local crossing minimization heuristics can be easily incorporated in the routing algorithm. After routing, there can be overlaps of edge segments and the minimum distance requirement is violated. The third phase performs mental map preserving layout adjustment [7, 10] to remove overlaps while minimizing the node movement.

The quality of the obtained drawings is mostly influenced by the node placement phase which is the main contribution of this paper. Edge routing and overlap removal phases will not be described any further since good solutions exist in the given references. In the node placement phase nodes are placed in a two-dimensional rectangular grid each node occupying one or more grid cells. Node placement is divided into two stages. In the first stage all nodes are treated to be of the same size occupying exactly one grid cell. In the second stage a node can take several grid cells proportional to its given size. The second stage could be used alone, but obtaining an initial approximation with the unit size nodes often results in better layouts.

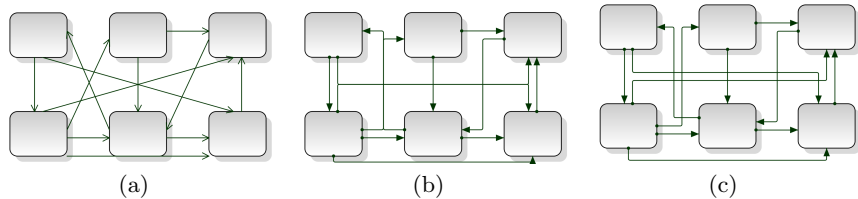

Fig. 1: Graph obtained after each of the 3 phases: **(a)** node placement; **(b)** edge routing; **(c)** normalization.

Similarly to [3], we formalize the node placement as an optimization problem to minimize the total edge length subject to constraints that no two nodes cover the same grid cell. The basis of our algorithm is inspired by the simulated annealing idea. We perform a greedy optimization that iteratively moves a nodes to a better place or swaps two nodes. This process is augmented with some random displacement. The idea of using a grid for node placement(although for straight-line drawings) together with simulated annealing is used in drawing of biochemical networks [12,13,15]. But in our algorithm we extend it with repeated global compaction steps that allow to escape from the many local minima of the optimization problem. Figure 2 shows a simple example where no node can be moved to improve the layout but the compaction step of our algorithm in vertical direction produces the optimal layout. In such small examples random displacements helps to find the optimum, but similar cases when groups of nodes have to be shifted often occur in larger graphs where randomization is too weak.

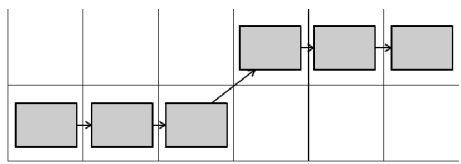

Fig. 2: Situation where compaction is needed to improve layout.

## 3 Detailed Description

In the input to the node placement algorithm we are given a graph $G = (V, E)$ with a node set $V$ and edge set $E$ to be laid out and the minimum width $w_i$ and height $h_i$ of each node. In output the algorithm gives the top-left corner $(x_i, y_i)$ of each node.

To deal with nodes of different sizes (relevant only in the second stage of the algorithm) we need to calculate the size of a grid cell. We assume the grid cell to be a square of side length $c$ which is calculated as

$$c = \begin{cases} L_{max} & \text{if } L_{max} < 3L_{min} \\ \frac{3L_{min}}{2} & \text{if } 3L_{min} \leq L_{max} < 15L_{min} \\ \frac{L_{max}}{30} & \text{if } 15L_{min} \leq L_{max} \end{cases}, \qquad (1)$$

where $L_{min} = \min(\min(w_i + \delta), \min(h_i + \delta))$ and $L_{max} = \max(\max(w_i + \delta), \max(h_i + \delta))$. The main case of $c$ is the middle one. The first case is chosen when all nodes are of a similar size and we define $c$ such that all boxes take only one grid cell for more pleasant results. The third case prevents excessive memory usage in case of widely different node sizes.

When nodes are placed in the grid, they are given integer coordinates and sizes. The top-left corner of a node $v_i$ in the grid will be denoted by $(x'_i, y'_i)$. Its width in the grid $w'_i$ is calculated as $\lceil \frac{w_i + \delta}{c} \rceil$. Its height in the grid $h'_i$ is calculated as $\lceil \frac{h_i + \delta}{c} \rceil$.

We can use different functions for the edge length to be minimized. Common examples include Euclidean or Manhattan distance. To deal with nodes of different sizes better, we use a distance function $d(v_i, v_j)$ between two nodes $v_i$ and $v_j$ defined as follows:

$$d(v_i, v_j) = d_e(v_i, v_j) + \frac{1}{20} \min\left(\frac{|x_i^c - x_j^c|}{w'_i + w'_j}, \frac{|y_i^c - y_j^c|}{h'_i + h'_j}\right), \qquad (2)$$

where $d_e(v_i, v_j)$ is the Euclidean distance between the node rectangle borders and $x_i^c = x'_i + 1/2 w'_i$ and $y_i^c = y'_i + 1/2 h'_i$ are center coordinates of the nodes. The second addend helps to align node centers when the distance between their borders is approximately equal. The constant $1/20$ was chosen experimentally to balance the need for short edges with alignment of node centers.

### 3.1 Compaction

An essential step of the proposed layout algorithm is compaction, which performs global layout improvements and simultaneously creates new empty places in the grid. We use a quadratic programming approach [7,10] where compaction in one dimension is expressed as minimization of a quadratic function subject to two-variable linear constraints. The function is constructed to minimize the total edge length but constraints keep the minimum distances between nodes and maintain their relative ordering.

Compaction is done separately in horizontal and vertical directions. Let us consider the horizontal direction; the vertical one is similar. The relative ordering is expressed as a visibility graph. A visibility graph is a directed graph with the same set of nodes $V$ but with a different set of edges $S$. There is a directed edge $(i, j) \in S$ in the visibility graph if and only if $x'_j > x'_i$ and it is possible to connect nodes $v_i$ and $v_j$ with a horizontal line segment without overlapping any other node. The visibility graph can be constructed with a sweep-line algorithm in time $|V| \log |V|$ but in our case we can extract it directly from the grid in time proportional to the number of grid cells.

We construct the following optimization problem

$$\text{minimize} \sum_{(i,j) \in E} (z_i + 1/2 w'_i - (z_j + 1/2 w'_j))^2 \qquad (3)$$
$$\text{subject to } z_j - z_i \geq d_{ij}, (i, j) \in S$$

where $d_{ij} = \gamma \cdot w'_i$ and $\gamma \geq 1$ is a coefficient that defines how much empty space will be left between nodes. To obtain the maximum compaction we should set $\gamma = 1$. Such setting is desirable at the final few iterations of the algorithm but otherwise using $\gamma > 1$ leaves some empty places between nodes giving additional freedom for node movement to find a better solution.

To perform compaction, the visibility graph is constructed from the current node positions, the optimization problem is constructed and its minimum is found by using the solver described in [10]. The node positions are calculated as $x'_i = \lfloor z_i \rfloor$. Since $w'_i$ are integer and $\gamma \geq 1$, the rounded values satisfy $x'_j - x'_i \geq w'_i$ and non-overlapping of nodes is ensured. The rounded solution may not be optimal with respect to the integer variables $x'_i$ but is good enough for our purposes. Note that compaction uses a quadratic edge length function but node swapping uses a linear distance defined by equation (2). In our case such mismatch does not create obvious bad effects since (2) is used when swapping nodes and compaction respects the obtained ordering via the constraint graph.

Compaction is used also to switch from the first stage of the algorithm where all nodes are of a unit size to the second stage with the real node sizes. The switch is done by simply compacting with $d_{ij}$ calculated from the new node sizes. An example of horizontal compaction with different $\gamma$ values is shown in Figure 3.

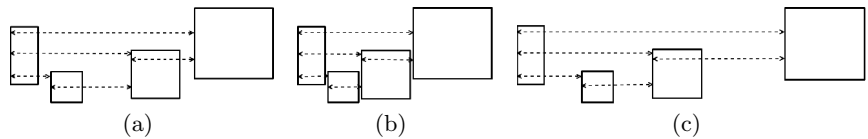

(a) (b) (c)

Fig. 3: An example graph of four nodes with their horizontal visibility graph(dashed). **(a)** before compaction; **(b)** after compaction with $\gamma = 1$; **(c)** after compaction with $\gamma = 2$.

## 3.2 Algorithm Pseudocode

The pseudocode for the node placement is shown in Algorithm 1. As the first steps, the grid of size $5\sqrt{|V|} \times 5\sqrt{|V|}$ is initialized and nodes are randomly placed in the grid. The first stage of the algorithm(lines 6-20) treats each node of size 1. The grid is dynamically expanded during layout, if required. The algorithm works in iterations and the number of iterations *iterationCount* is taken proportional to $\sqrt{|V|}$. At each iteration, local optimization is performed that decreases the total edge length. The optimization process is based on the simulated annealing idea. It requires the notion of temperature $T$ which influences how much node positions are perturbed by randomness. The starting temperature $T$ is set equal to $2\sqrt{|V|}$ to allow nodes to be placed almost everywhere initially. The temperature is smoothly reduced by a cooling coefficient $k$ until it reaches the lowest temperature $T_{min}$; we take $T_{min}$ equal to 0.2. The cooling coefficient $k$ is calculated in line 5 such that $T$ reaches $T_{min}$ in *iterationCount* iterations.

To perform local optimization, every node is moved to a location that minimizes the total length of its adjacent edges. We use a heuristic to calculate this location approximately. Calculating the optimal location is expensive and actually is not needed since the added random displacement disturbs it anyway. We calculate an initial estimate to node's position $(x,y)$ that minimizes the Manhattan distance to the adjacent nodes. Such point is found as a median of the neighbors' centers. A random displacement proportional to the temperature $T$ is added to that point.

Then we search the closest place to $(x,y)$, where $v_j$ can be put (line 10). We calculate the Manhattan distance $d$ of the closest free place to $(x,y)$. Then we check all cells within Manhattan distance $d+1$ from $(x,y)$ and choose the position with the least total edge length according to (2) to place $v_j$. If this place is different from the location of $v_j$ from the previous iteration, we leave the node there. Otherwise, we try to swap it with the nodes nearby. We do this by checking the nodes residing in adjacent grid cells to $v_j$. For each of these we calculate the gain of the total edge length if we swap the adjacent node with $v_j$. If the gain is positive we swap the nodes.

Compaction is performed every 9-th iteration each time changing direction(lines 15-18). The variable *compactionDir* defines direction in which compaction is be performed, *true* for horizontal direction *false* for vertical. Compaction is performed by function compact(boolean *horDirection*, float $\gamma$, boolean *expand*) described in section 3.1. The parameter *expand* is true if boxes need to be expanded to it's real sizes, otherwise it is false. In line 16 compaction is done with $\gamma = 3$ and direction is changed after every compaction (line 17). The temperature $T$ is reduced at the end of the iteration (line 19).

In lines 21 and 22 switching from the first stage to the second is done by performing compaction with the new node sizes. The second stage (lines 23-37) is similar to the first one, only all boxes are treated with their prescribed sizes. Searching for a place for a node has to check if all grid cells under a larger node are free. This modification influences node swapping – there may be cases when adjacent nodes of different sizes cannot be swapped. This is the main motivation why the first stage with unit node sizes is beneficial. In line 33 compaction is done with gradually decreasing $\gamma$ which becomes 1 in the last 3 compactions. In this way the available free space for node movement is gradually reduced giving more emphasis to node swapping.

```
1  Initialize the grid of size 5√|V| × 5√|V| Put nodes randomly into grid (treat
   them all 1 × 1 sized);
2  compactionDir=true;
3  iterationCount=90√|V|;
4  T=2√|V| ;
5  k=(0.2/T)^(1/iterationCount);
6  for ( i=0; i<iterationCount/2; i++ ) do
7      for (j=1; j ≤ |V|; j++) do
8          x=neighboursMedianX(v_j) + random(−T, T);
9          y=neighboursMedianY(v_j) + random(−T, T);
10         Put v_j near (x,y);
11         if v_j has not changed it's place from the previous iteration then
12             Try to swap v_j with nodes nearby;
13         end
14     end
15     if iterationCount mod 9 == 0 then
16         compact(compactionDir, 3, false);
17         compactionDir=!compactionDir;
18     end
19     T=T·k;
20 end
21 compact(true, 3, true);
22 compact(false, 3, true);
23 for ( i=iterationCount/2+1; i<iterationCount; i++ ) do
24     for (j=1; j ≤ |V|; j++) do
25         x=neighboursMedianX(v_j) + random(−T·w'_j, T·w'_j);
26         y=neighboursMedianY(v_j) + random(−T·h'_j, T·h'_j);
27         Put v_j near (x,y);
28         if v_j has not changed it's place from the previous iteration then
29             Try to swap v_j with nodes nearby;
30         end
31     end
32     if iterationCount mod 9 == 0 then
33         compact( compactionDir, max(1, 1 + 2(iterationCount−i−30)/(0.5iterationCount)), false) ;
34         compactionDir=!compactionDir;
35     end
36     T=T·k;
37 end
```

**Algorithm 1:** Main algorithm

### 3.3 Other Starting Layouts

The algorithm described above starts with assigning random positions to the nodes and later it needs relatively many iterations to find a good layout. We consider two other initial layouts for increased speed or quality – force directed placement and arrangement by breadth first search(BFS). The constants of the algorithm have to be adjusted depending on the chosen initial layout.

One possibility is to use force directed placement for the initial positions of nodes. Fast methods [11] are known for the force directed placement. Node coordinates in the grid are initialized from the rounded results of the force directed placement. Since force-directed placement gives a good approximation to the minimum edge length, *iterationCount* can be set constant and we made it equal to 100. Compaction is done at every 3-rd iteration. The starting temperature should be small, we set $T = 3$.

Another possibility for the starting layout is to use incremental placement where nodes are added one by one. We execute a breadth-first search starting from some arbitrary chosen node and add nodes to the grid in this order. The position for each node is chosen as a free place that minimizes the total distance to already placed nodes. We found that BFS placement gives good results with small graphs or graphs with a small degree. For this starting layout we chose coefficients as follows: $iterationCount = 10\sqrt{|V|}$, $T = 0.2\sqrt{|V|}$. Compaction is done after every 3-rd iteration.

## 4 Results

We have tested our algorithm on many artificial and real world graphs. Figure 8 shows some examples. The algorithm produces pleasant drawings with small area and low number of crossings and edge bends. For small graphs like these, all three proposed initial placement methods produce similar results.

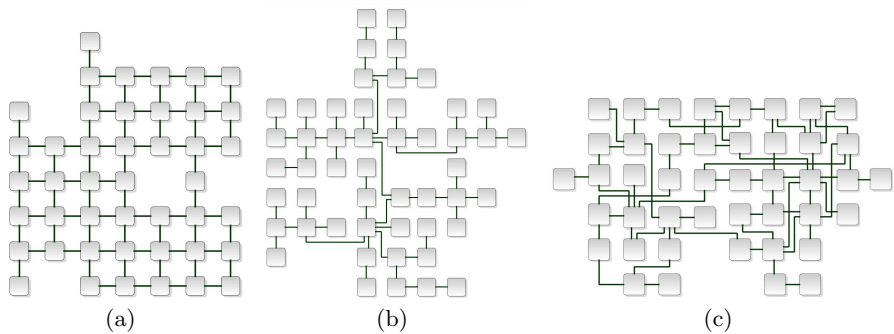

(a) (b) (c)

Fig. 4: Examples of the tested graphs. (a) partial grid graph; (b) tree graph; (c) random graph.

To test the quality and performance of our algorithm we run it on three automatically generated graph classes – partial grids, random trees and random graphs, see Figure 4. Three modifications of the algorithm with different initial placements were tested for each class, the time for initialization is included in the measurement. For each class of graphs random instances were generated of progressively increasing size, 10 instances for each size. The running time and the average number of crossings per edge were calculated. To be independent of the routing algorithm, edges were treated as straight line segments connecting the node centers for the crossing calculation. The results were averaged over the 10 generated instances.

The running time measurements are similar for all three graph classes. The running time mostly depends on the number of iterations chosen in each case of initialization, random case being the slowest and force-directed case the fastest. The measurements indicate that the time for initial placement does not add much overhead.

The results for the partial grid graphs are shown in Figure 5. A partial grid graph is a square grid with the specified number of nodes where 10% of nodes are randomly removed. The results show that the algorithm with all the initial placement methods produce a planar layout of small instances (up to about 1000 nodes) but further only force-directed initialization is able to recover the graph structure correctly, BFS initialization being the worst. It has to be mentioned that, if we increase the number of iterations of the BFS case to match the random case, we obtain drawings of similar quality. But our intention for the BFS method was to check whether we can improve running time with a better initialization. Tests showed that BFS initialization does not give any advantage over the random one.

The quality on tree graphs is similar for all three modifications (Figure 6). None is able to produce completely planar drawings of larger instances, although the crossing count is small. The BFS method has slightly more crossings than the other two.

Random graphs are generated by including randomly chosen node pairs as edges in the graph with density $|E| = 1.2|V|$. The quality on random graphs is similar for all three methods (Figure 7). That is expected since random graphs cannot be drawn with significantly less crossings than any of these methods produce.

Overall, the best initialization method is force-directed, which produce the best drawing quality in the least running time. Of course, this option depends on the quality and performance of the available force-directed placement implementation.

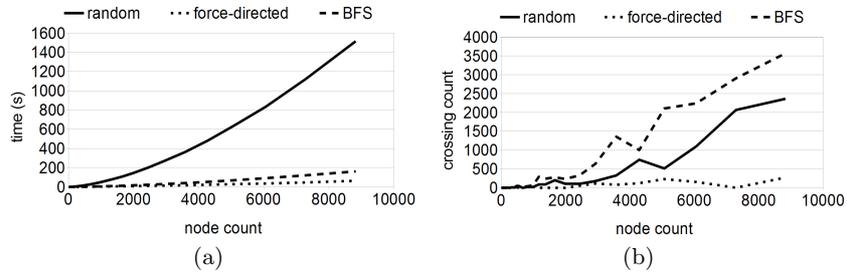

Fig. 5: The running time and crossing count depending on node count of partial grid graphs. (a) running time; (b) crossing count.

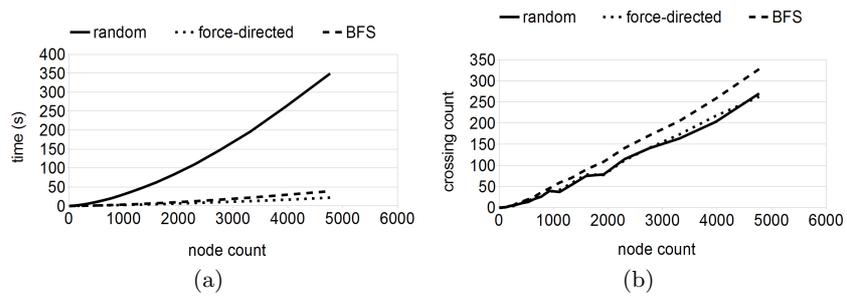

Fig. 6: The running time and crossing count depending on node count of tree graphs. (a) running time; (b) crossing count.

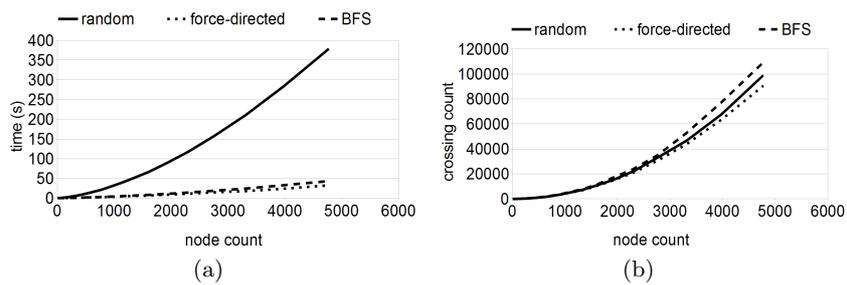

Fig. 7: The running time and crossing count depending on node count of random graphs. (a) running time; (b) crossing count.

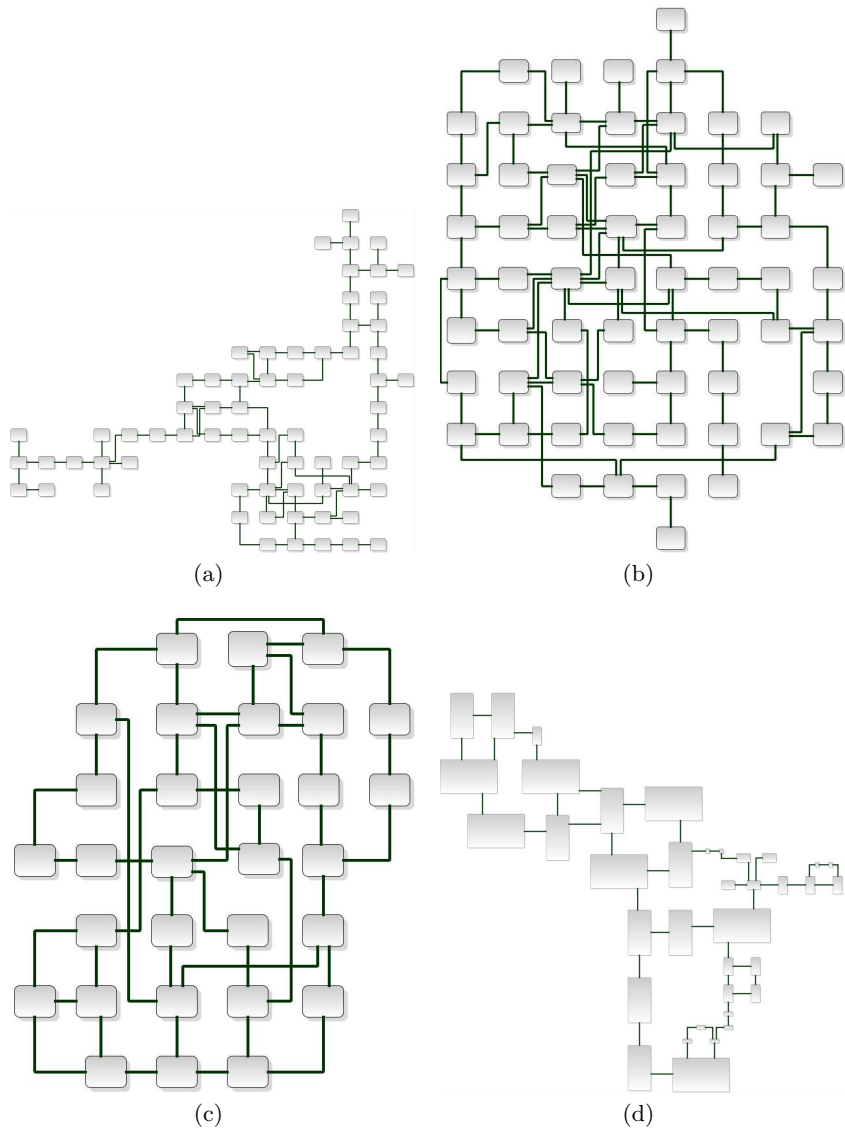

Fig. 8: Examples of layouts produced with the proposed algorithm. (a) the graph presented in [13]; (b) the graph presented in [6]; (c) the graph presented in [4]; (d) a graph with nodes of different sizes.

## References

1. Biedl, T., Kant, G.: A better heuristic for orthogonal graph drawings. Computational Geometry 9(3), 159–180 (1998)
2. Biedl, T.C., Kaufmann, M.: Area-efficient static and incremental graph drawings. In: AlgorithmsESA'97. pp. 37–52. Springer (1997)
3. Biedl, T.C., Madden, B.P., Tollis, I.G.: The three-phase method: A unified approach to orthogonal graph drawing. In: Graph Drawing. pp. 391–402. Springer (1997)
4. Bridgeman, S., Fanto, J., Garg, A., Tamassia, R., Vismara, L.: Interactivegiotto: An algorithm for interactive orthogonal graph drawing. In: Graph Drawing. vol. 1353, pp. 303–308. Springer (1997)
5. Di Battista, G., Didimo, W., Patrignani, M., Pizzonia, M.: Orthogonal and quasi-upward drawings with vertices of prescribed size. In: Graph Drawing. pp. 297–310. Springer (1999)
6. Di Battista, G., Garg, A., Liotta, G., Tamassia, R., Tassinari, E., Vargiu, F.: An experimental comparison of four graph drawing algorithms. Computational Geometry 7(5), 303–325 (1997)
7. Dwyer, T., Marriott, K., Stuckey, P.J.: Fast node overlap removal. In: Graph Drawing. pp. 153–164. Springer (2006)
8. Fößmeier, U., Heß, C., Kaufmann, M.: On improving orthogonal drawings: The 4m-algorithm. In: Graph Drawing. pp. 125–137. Springer (1998)
9. Fößmeier, U., Kaufmann, M.: Drawing high degree graphs with low bend numbers. In: Graph Drawing. pp. 254–266. Springer (1996)
10. Freivalds, K., Kikusts, P.: Optimum layout adjustment supporting ordering constraints in graph-like diagram drawing. In: Proceedings of Latvian Academy of Sciences, Section B. pp. 43–51. No. 1 (2001)
11. Hachul, S., Jünger, M.: An experimental comparison of fast algorithms for drawing general large graphs. In: Graph Drawing. pp. 235–250. Springer (2006)
12. Kojima, K., Nagasaki, M., Miyano, S.: Fast grid layout algorithm for biological networks with sweep calculation. Bioinformatics 24(12), 1433–1441 (2008)
13. Kojima, K., Nagasaki, M., Miyano, S.: An efficient biological pathway layout algorithm combining grid-layout and spring embedder for complicated cellular location information. BMC Bioinformatics 11, 335 (2010)
14. Lengauer, T.: Combinatorial algorithms for integrated circuit layout. John Wiley & Sons, Inc. (1990)
15. Li, W., Kurata, H.: A grid layout algorithm for automatic drawing of biochemical networks. Bioinformatics 21(9), 2036–2042 (2005)
16. Six, J.M., Kakoulis, K.G., Tollis, I.G., et al.: Techniques for the refinement of orthogonal graph drawings. J. Graph Algorithms Appl. 4(3), 75–103 (2000)
17. Tamassia, R.: On embedding a graph in the grid with the minimum number of bends. SIAM Journal on Computing 16(3), 421–444 (1987)